\documentclass[onecollarge]{svjour2}       

\journalname{General Relativity and Gravitation}

\begin{document}

\title{Causality and the speed of sound }

\author{George F.R. Ellis, Roy~Maartens, and Malcolm A.H. MacCallum}

\institute{George F.R. Ellis \at
               Mathematics and Applied
Mathematics Department,
University of Cape Town, Cape~Town~8001, South Africa\\
             \email{ellis@maths.uct.ac.za}          \\
           \and Roy~Maartens
           \at Institute of Cosmology \& Gravitation, University of
Portsmouth, Portsmouth~PO1~2EG, UK\\
             \email{roy.maartens@port.ac.uk}          \\
 \and Malcolm A.H. MacCallum
 \at School of Mathematical Sciences, Queen Mary, University of
London, London~E1~4NS, UK\\
             \email{m.a.h.maccallum@qmul.ac.uk}}

\date{Received: date / Accepted: date}

\maketitle

\begin{abstract}

A usual causal requirement on a viable theory of matter is that
the speed of sound be at most the speed of light. In view of
various recent papers querying this limit, the question is
revisited here. We point to various issues confronting theories
that violate the usual constraint.


\end{abstract}

\section{ Introduction}

In cosmology and astrophysics, it is usually assumed that the
speed of sound $c_s$ cannot exceed the speed of light $c$; indeed
$c_s> c$ is taken as a criterion for rejecting theories. This is
the view put in established texts (see, e.g.,
Refs.~\cite{hawell73,wald}), in earlier papers (see e.g.\ \cite{Mon80}),
and in more recent work. For example,
it has been argued that accelerating k-essence models in cosmology
are ruled out because the scalar field fluctuations in this case
must propagate superluminally~\cite{bonetal06}. Low-energy
effective field theories have been rejected when, despite having
Lorentz-invariant Lagrangians, they admit superluminal
fluctuations~\cite{Adams:2006sv}. Subluminal propagation of field
fluctuations has been imposed as a condition on a relativistic
gravitation theory for the MOND
paradigm~\cite{Bekenstein:2004ne,Bekenstein:2007iq}.

Other recent papers however challenge these standard views of
causality (see, e.g.,
Refs.~\cite{brun06,babmuk05,Mukhanov:2005bu,ericetal02}), and
propose matter models that allow superluminal signal propagation,
which can lead to interesting effects in cosmological and
astrophysical contexts. (These are not the first such models: see e.g.\
\cite{BluRud68,Rud68}.)

Here there is a clash of cultures between an approach where a
matter model, usually based on an assumed Lagrangian, can be
chosen freely, and an approach where fundamental relativistic
principles constrain the matter model. The immediate problem with
matter models that have superluminal physical modes is that they
may violate causality, one of the most basic principles of special
relativity theory. To be specific, many proposals violate
``Postulate (a): Local Causality'', given on page 60 of
Ref.~\cite{hawell73}. This is usually taken as a criterion for
rejecting a theory, because local causality is taken to be an
absolute requirement on all theories:\vspace*{\baselineskip}

\noindent\textbf{Comment 1}: \emph{The strictly relativistic
position is that matter models which violate the causality
requirement $c_s^2 \leq c^2$ are ruled out as
unphysical.}\vspace*{\baselineskip}

However there are a number of papers appearing in the current
literature that do not take this view. Instead, they drop the
usual speed of sound limit on fluid or scalar field models, but
maintain special relativity properties locally. To test if the
conservative view (as stated in Comment~1) is too restrictive, we
look here at some issues that arise from this drastic step. Thus
we consider foundational issues which have to be taken into
account in any proposal allowing such apparent causality
violations. We find that one can indeed construct a macroscopic
phenomenological theory that is covariantly well-defined and
respects standard principles including Lorentz invariance.
However, in such an approach, fundamental features in all of
physics (special relativity, electromagnetism, gravitation,
quantum field theory, etc.) are determined or affected by an
arbitrary matter model. Furthermore, it appears that such matter
models cannot be based in a relativistically acceptable
microphysical theory.

\section{ Macroscopic fluid models and the sound cone}

Consider a perfect fluid in a Special or General Relativity
context, with equation of state $p/c^2=w\rho$ relating the
pressure $p$ to the energy density $\rho$. If $w$ is constant, the
speed of sound is given by
\begin{equation}\label{speed}
{c_s^2 \over c^2} = {1\over c^2}{dp \over d\rho} = w\,,
\end{equation}
and if $w$ is slowly varying, this is still a good approximation.
Thus one can get $c_s^2>c^2$ easily: simply set $w >1$ in the
macroscopic description, i.e., presume that $p/c^2 > \rho > 0$.
Then the speed of sound cones lie outside the speed of light cones
in all directions\footnote{A perfect fluid is isotropic about its
fundamental velocity $u^\mu$, so that relative to this velocity
the speed of sound is the same in all directions.} at all events,
and fluid waves can propagate at speeds up to and including this
superluminal speed of sound. Of course this is far from ordinary
matter. It does not accord with anything so far experienced in the
real world. But does it cause serious problems in terms of causal
violations or Lorentz invariance, considered macroscopically?

Recently it has been claimed (e.g.,
Refs.~\cite{brun06,babmuk05,Mukhanov:2005bu,ericetal02}) that
there need not be problems with either issue, if one approaches
the problem in an open-minded way. The key point is the following:
we usually associate causality with the light cone, but any
suitable set of cones can be used to define the macroscopic
limiting speed, including sound cones that lie outside the light
cones. Assuming that (1)~the 4-velocity of the fluid defining the
sound cone stays inside the light cone, and (2)~photons and
gravitons still move at the speed of light, then the light cones
will always stay inside the sound cones and so will not give any
acausal propagation in the sense defined by the sound cones. The
basic essence of the relativistic causality requirement would be
preserved, though causal limits would be determined by the speed
of sound in the fluid rather than the speed of light.

Now, Lorentz invariance is broken by the sound cones in this case,
because a change of the observer's velocity will result in
apparently different speeds of sound in different directions
(unlike the case of light, which has the same speed in all
directions for all observers). This proposal might therefore
appear to violate Einstein's basic principle that the laws of
physics should be the same in all inertial frames. But we are
concerned with solutions of the basic equations, rather than the
symmetries of the equations themselves. All realistic fluid
solutions break Lorentz invariance\footnote{The fluid stress
tensor is only Lorentz invariant if $\rho + p/c^2 = 0$, but this
is just the degenerate case of a cosmological constant. It is not
a realistic fluid, and in any case cannot support matter
perturbations.}, in particular because of the uniquely
defined 4-velocity of the matter, and these solutions are no
exception. Equations can have an invariance not shared by their
solutions.

Indeed these papers are not considering general superluminal
motions, but superluminal signals referred to a specific rest
frame. These signals are just as non-Lorentz-invariant as sound
waves in normal fluids, which are subluminal with respect to the
rest frame of the fluid. A Lorentz transformation which preserves
the light cone will not in general preserve either subluminal or
superluminal signals of this sort, because it will not preserve
the 4-vector on which the sound cone is based. However, the rest
frame of the fluid and the sound velocity in it may be physically
well-defined, just as the rest frame in cosmology is defined by
the 4-velocity of the substratum. No physical violation is
involved in this aspect of the proposal. Lorentz-invariant
theories not only can have, but to model some aspects of reality
must have, non-Lorentz-invariant solutions (otherwise normal sound
waves would not be allowed either). The invariance then maps one
solution to another different one, rather than to itself.

In other observers' rest frames, in this case, causal limits will
again be determined by the speed of sound cone of the fluid rather
than the light cone. There is no way to send a signal into one's
past provided no signal, and no observer, travels outside the
sound cone, so this cone is itself the causal limit cone. The
argument is exactly the same as usual causality, just with
different cones: the cause of a here-and-now event must lie inside
the past sound cone and this cannot be reached from the future
sound cone. The cones are fixed and the same for all observers.
Thus causal paradoxes do not arise from this effect, because these
cones are fixed by the invariant fluid velocity; they are not
arbitrarily assignable by changing the observer's velocity.

{}From the viewpoint of standard relativity theory, a major
problem with this model is that {\em causality depends on the
matter content of the universe.} In the usual view, the same
causality applies to all physics, independent of the kind of
matter present; the causal limit has a much more fundamental
character. In the proposal discussed here, the notion of a causal
limit becomes arbitrarily dependent on the matter model, unlike
the relativistic proposal where it is the same for all matter.
Furthermore, according to the simplistic model discussed here,
arbitrary fluids could give arbitrarily large values for $c_s$;
there is no longer any effective limiting speed at macroscopic
scales.\vspace*{\baselineskip}

\noindent\textbf{Comment 2}: \emph{Lorentz invariance \emph{per
se} does not prohibit macroscopic theories with superluminal
sound: the speed of sound could limit causality rather than the
speed of light doing so. However this does not give a good theory
of relativistic type since $(a)$~causality depends on the matter
model, and $(b)$~there is then no upper limiting causal speed: any
speed is apparently possible}.\vspace*{\baselineskip}

One can reconcile these ideas only if there is some unique field that
defines causality, and somehow all other fields are prevented from
having a faster wave speed -- i.e., a very special kind of matter or
field exists that somehow has a fundamental role to play in all of
physics. In the standard view, that role is played by the metric
tensor. If there is another such field, it should be
identifiable.
Below we return to the issue of whether one can meaningfully
define other metric tensors associated with the new proposal given
here for limiting speeds.

\section{ Scalar fields and variational principles}

A fluid description is of course a highly simplified effective
theory, and one can propose more fundamental theories for the
matter. The most common one in cosmology is a scalar field. If the
Lagrangian density is ${\cal L}$, then the pressure and energy
density (in the natural frame, $u_\mu\propto
\partial_\mu\varphi$) are given by~\cite{Garriga:1999vw}
\begin{equation}\label{pr}
p=c^2{\cal L}\,,~ \rho=2X{\cal L}_{,X}-{\cal L}~\mbox{where}~
X\equiv {1\over 2}g^{\mu\nu}\partial_\mu\varphi
\partial_\nu \varphi\,.
\end{equation}
Fluctuations of the scalar field propagate with effective speed of
sound~\cite{Garriga:1999vw}
\begin{equation}\label{cs}
{c_s^2}={ p_{,X}\over \rho_{,X}}\,.
\end{equation}

A standard scalar field has Lagrangian density
\begin{equation}\label{lag}
{\cal L}=-X -V(\varphi)\,,
\end{equation}
so that
\begin{equation}\label{w}
p=c^2(-X-V)\,,~\rho=-X+V\,.
\end{equation}
It follows from Eq.~(\ref{cs}) that
\begin{equation}
c_s^2=c^2\,,
\end{equation}
independently of the scalar field's potential. This can be
explained from the local special relativistic viewpoint by the
fact that for high frequency waves only the
$g^{\mu\nu}\nabla_\mu\nabla_\nu \delta \varphi$ terms in the wave
equation for scalar field fluctuations are significant.
Alternatively, using wave-particle duality, massive particles can
move with any four-velocity inside the light cone, so that the
limiting speed is $c$.

In the cosmological context, Eq.~(\ref{w}) and the Klein-Gordon
equation imply
\begin{equation}
{\dot p \over \dot \rho}=1+{2V_{,\varphi} \over 3H\dot\varphi}\,.
\end{equation}
Clearly, $\dot p/\dot\rho\neq c_s^2/c^2$. Indeed, even in
slow-roll inflation, when $\dot p/\dot \rho\approx -1$, we have
$c_s^2=c^2$. This difference between $\dot p/\dot \rho$ and
$c_s^2/c^2$, unlike a perfect fluid with $\dot
p/\dot\rho=w=c_s^2/c^2=\,$constant, reflects the presence of
intrinsic entropy perturbations in the field~\cite{ent}.

Scalar fields with generalized, non-standard Lagrangians allow
superluminal speeds of sound. For example, if we generalize
Eq.~(\ref{lag}) to
\begin{equation}
{\cal L}=-F(X)-V(\varphi)\,,
\end{equation}
then Eqs.~(\ref{pr}) and (\ref{cs}) give
\begin{equation}
{c_s^2\over c^2}={F_{,X} \over F_{,X}+2XF_{,XX}}\,.
\end{equation}
Thus $c_s^2>c^2$ is possible for appropriate choices of
non-standard kinetic term $F(X)$. Examples include
nonlinear complex scalar fields \cite{BluRud68}, k-essence
models of inflation and dark
energy~\cite{Mukhanov:2005bu,ericetal02}, and Born-Infeld type
models of scalar fields that supposedly can transmit information
from inside a black hole~\cite{babmuk05}.

The logic of these proposals is that any Lorentz-invariant
Lagrangian leads to acceptable models. By contrast, if we do not
give primacy to ad hoc matter models, but instead impose
relativistic principles as
fundamental~\cite{bonetal06,Adams:2006sv,Bekenstein:2004ne}, then
the Lagrangian is ruled out as non-physical, since such scalar
fields violate the most basic principles of special relativity
theory. \vspace*{\baselineskip}

\noindent\textbf{Comment 3}: \emph{Existence of a variational
principle does not necessarily imply existence of corresponding
matter. If the solutions violate causality, this is \emph{a
priori} a good reason to believe that the variational principle is
unphysical.} \vspace*{\baselineskip}

Non-standard kinetic terms are highly problematic. There is no
experimental evidence for them, and no compelling reason to think
they are physical. As in the previous case, the notion of a causal
limit becomes arbitrarily dependent on the matter model. As
arbitrary scalar fields can give arbitrarily large values for the
speed of sound, there is again no longer an effective limiting
speed at macroscopic scales. Furthermore, the speed of sound can
change in space or time from subluminal to superluminal or vice versa,
as in k-essence models \cite{bonetal06}. All these seem problematic from a
relativistic viewpoint.

\section{ Alternative metrics}

The sound cones for any given fluid or scalar field can be
represented by an appropriate metric tensor of hyperbolic type. If
$u^\mu$ is the matter 4-velocity, and $h_{\mu\nu}=g_{\mu\nu}+u_\mu
u_\nu$ projects into the rest space at each event, then one can
define the metric
\begin{equation}\label{metric}
(G^{-1})^{\mu\nu} = g^{\mu\nu} -(c^2-c_s^2) h^{\mu\nu}\,,
\end{equation}
which gives the characteristic cones, and the rays are given by
\begin{equation}\label{metric1}
G_{\mu\nu} = g_{\mu\nu} +\frac{c^2}{c_s^2}(c^2-c_s^2)h_{\mu\nu}\,.
\end{equation}
Thus the sound cones are given by this metric, and it is useful
for visualization purposes to draw them; they will lie outside the
light cones when $c_s^2>c^2$. One can then rephrase the point by
saying that there are two metrics: in the case of superluminal
sound those metrics agree that the interior of the light cone
consists of timelike vectors.

However, some of those who argue that there is no problem with
causality go much further: they say that if there is a
superluminal mode, one can just re-define the physical metric to
be the sound metric, Eq.~(\ref{metric1}), based on the
pathological wave equation, and then the problem disappears. For
example, Ref.~\cite{brun06} encapsulates this view by stating
that: ``causality should not be expressed in terms of the
chronology induced by the gravitational field \dots there is no
clear reason why a metric or chronology should be preferred to the
other \dots the gravitational metric field is just one particular
field on spacetime and there is no clear reason why it should be
favored''.

We profoundly disagree. The spacetime metric is special, despite
these claims: it determines time measurements and spatial
distances, as well as the free-fall motion that is the essential
basis of the equivalence principle, and hence the basis for
identification of gravity as being expressed through space-time
curvature~\cite{hawell73,wald}. It does so alike for all matter
and fields. Furthermore, if we abandon the spacetime metric as
arbiter of causality, then we could find that
the sound metric was in
some places superluminal and in others subluminal, so that,
at least at some
events and in some directions, part of the light cone could lie
outside the sound cone. Then photons and gravitons could propagate
acausally relative to the wave-equation metric (i.e., the one that
``makes" the superluminal modes causal).\vspace*{\baselineskip}

\noindent\textbf{Comment 4}: \emph{Just defining something as a
metric does not mean that it has all the properties of the
preferred spacetime metric. The spacetime metric is preferred in
terms of clock measurements and free fall $($geodesic$)$ motion
$($including light rays$)$, thus underlying General Relativity's
central theme of gravity being encoded in spacetime curvature. It
is also related to the Lorentz group that underlies all particle
physics, and hence to the limiting speed of motion of all
particles following from the special relativistic equations of
motion at each point}.\vspace*{\baselineskip}

The speed of sound metric simply does not have all these
properties. It is in no way equivalent to the spacetime metric.
And because of the relation to the Lorentz group and hence to
special relativity, it is the speed of light cones that will
represent correctly the microscopic limiting speed.

\section{ Microscopic fluid models}

The spacetime metric defines the Lorentz transformations that
underlie microscopic physics, and indeed is the basis of the
definitions of variables that occur in current fundamental
theories. The resulting special relativistic microscopic equations
of motion prevent any real physical particle or associated signal
moving faster than light; relativity only allows particles
travelling at up to the speed of light (they cannot be accelerated
to greater speeds because of the unbounded increase of the
relativistic inertial mass). And here is the real problem for any
macroscopic theory of superluminal signal propagation: you cannot
base it in a microscopic theory of matter consistent with special
relativity theory, because there is no underlying microscopic
mechanism that could support such macroscopic behaviour. In
essence: you cannot have macroscopic signals propagating at a high
speed on the basis of particles and fields all of which travel at a
slower speed.

As stated in Ref.~\cite{bonetal06} in relation to scalar fields:
``The idea is of course that $\varphi$ is an effective low energy
degree of freedom of some fundamental high energy theory which
should satisfy basic criteria: among them, most importantly,
Lorentz invariance and causality. No information should propagate
faster than the speed of light $c=1$.'' Now if this is true of the
microscopic theory underlying the macroscopic theory, it has to be
true of the macroscopic theory as well. Any superluminal signals
would have to be mediated by particles travelling faster than
light, but such particles do not exist. More generally, the
low-energy effective theory, if it is to have a high-energy
completion that satisfies the basic postulates of quantum field
theory, will not admit superluminal signals~\cite{Adams:2006sv}.

There are circumstances in which microscopic theory supports
effects which can be apparently superluminal.
Some of these arise within quantum field theory.\footnote{Although
  Feynman diagrams can be thought of as
including particles travelling backwards in time, we do not accept
the interpretation that one can therefore send macroscopic
physical signals backwards in time via such effects or via virtual
particles.} Quantum entanglement
effects can link spacelike-separated events and in that sense work
superluminally -- but such effects depend on careful state
preparation and are so fragile that they cannot be causally
effective on macroscopic scales. They seem unable to carry
arbitrarily chosen information faster than light, which a genuine
wave is able to do.
For the matter model discussed in \cite{BluRud68} the superluminality is
identified in \cite{Rud68} as originating in
renormalization of a negative bare mass and gives rise to kinetic
energies with no lower bound (and hence without a well-defined ground
state). In
\cite{LibSonVis02} the superluminal velocities which are found in the
Scharnhorst effect (where they are due to QED corrections in the
Casimir effect) are argued not to lead to causality violations because they
define a causal cone in the frame of the Casimir plates which can used
to redefine causality in the same way as the sound cones discussed
above; however, they exist only locally between the plates and
therefore cannot be used to signal.
It is also possible to construct lattice models with
$p> \rho c^2$ and $dp/ d \rho > c^2$, but the sound speed (i.e.\ the
signal propagation speed) remains subluminal~\cite{Caporaso:1979uk}.

There are also effects arising in the case of light itself from
quantum mechanical tunnelling. These have been observed and it has been
argued \cite{CarMig03} that they are essentially the same as the
effects seen for ``X-shaped'' light beams, although the latter can be
described by classical field
theory (see e.g.\ \cite{Rec01}). These two effects have generated an
extensive literature. However, strong counterarguments have been given
against both the accuracy of some of the experiments and their theoretical
interpretation \cite{BigHag01,RinMea01}: in particular the X-shaped
beam effects have been
argued to be due to a ``scissors effect'' where the point at which two
beams interfere constructively moves at superluminal speed
\cite{RodThoXav01}, an effect similar to the possible ``superluminal''
movement of the end of a beam from a lighthouse.  Even
those active in working on these waves do not unambiguously argue
that they lead to superluminal propagation of information or signals
\cite{Rec01,RanFabPaz06}.

The X-shaped beams can be discussed in terms of a dispersion relation,
and one then has the well-known issues of distinguishing the phase
velocity, the group velocity, and the signal velocity: dispersion is
in fact common in the acoustics of media less simple than the models
discussed above. For example, in the X-shaped beams, the peak may be
travelling at a group velocity, apparently superluminally, although
the leading edge travels at the phase speed $c$ -- but this is only
possible until the
peak catches up with the front \cite{RodThoXav01}.
One can have superluminal group velocities, but this does not lead,
in any physically plausible example we know of, to speeds of sound
greater than those of light (see e.g. Ref.~\cite{Shore:2003jx})\footnote{It
appears that one can also analyze this situation, and that of entanglement,
using the information-theoretic interpretation of entropy, and again reach
the conclusion that superluminal propagation is not possible. We
thank G. Thompson for raising this point.}.

Thus it appears that
none of these examples provides a convincing contradiction to the
principle that  if the microscopic theory has a limiting
speed, that of light, then so does any macroscopic theory based on
it. You can send signals at the speed of light, characterized by
the light cones in the usual way, but no faster. Thus causality
for all particles is associated with these specific cones. 
Although you can create macroscopic descriptions of material with
superluminal signals, as discussed in the preceding sections, you
cannot adequately base them in a microscopic theory that obeys
fundamental requirements of theoretical physics. Indeed the
limiting state one can conceivably get from a viable microscopic
description is apparently ``stiff matter'' as proposed by
Zeldovich~\cite{zel62}, with equation of state $p/c^2 = \rho$, so
that $c_s^2 = c^2$. \vspace*{\baselineskip}

\noindent\textbf{Comment 5}: \emph{It appears that none of the
above causality-violating theories can be based on a microscopic
matter model that obeys special relativity principles. A viable
theory of causal limits must be a consistent whole for microscopic
and macroscopic physics. The extremely well-tested theory of
special relativity then insists on the speed of light as the local
limiting speed of causality.}

\section{ Conclusion}

As in the case of varying speed of light theories (see, e.g.,
Ref.~\cite{ell07} for a discussion), one must take physics as a
whole into account whenever proposing theories of superluminal
speed of sound; one cannot just tinker with some part of physics
without thinking of the consequences for the whole. Special
relativity is one of the best tested theories we have. It is not
enough to put forward \emph{ad hoc} matter models that violate its
principles and effectively alter all of physics. In order to make
a serious challenge of this nature, one needs a solid
justification, and a really plausible reason to abandon it --
backed up (in due course) by experimental data.

We have strongly made the case for a particular viewpoint based on
an understanding of present day physics. However it is always
possible that this understanding could be contradicted by
experiment and observation.\footnote{For example, it is in
principle possible to test the speed of sound of dark energy in
the universe by observations of supernova luminosity distances,
weak lensing and anisotropies in the cosmic microwave background.}
In that case, theory must give way to data.

\section*{Acknowledgements}

We are grateful to B. Bassett, N. Bilic, J.M. Charap, A. Feinstein, L. Herrera,
J. Ovalle, A. Polnarev and R.H. Sanders for comments which
enabled us to improve an earlier draft of this paper and/or add
pertinent references.

\end{document}